\documentclass[journal]{IEEEtran}
\usepackage{amsmath,amsfonts}
\usepackage{array}
\usepackage[caption=false,font=scriptsize,labelfont=sf,textfont=sf]{subfig}
\usepackage{textcomp}
\usepackage{stfloats}
\usepackage{url}
\usepackage{verbatim}
\usepackage{graphicx}
\usepackage{tabularx}
\usepackage{cite}
\usepackage{xcolor}
\usepackage{soul}
\usepackage{multirow}
\usepackage{mathtools}
\usepackage{algorithm}
\usepackage{algpseudocode}

\begin{document}


\title{Deep Learning Methods for IoT Device Authentication Using Symbols Density Trace Plot}

\author{
\IEEEauthorblockN{Da Huang,~\IEEEmembership{Graduate Student Member, IEEE}, Akram Al-Hourani,~\IEEEmembership{Senior Member, IEEE}, \\Kandeepan Sithamparanathan,~\IEEEmembership{Senior Member, IEEE}, and Wayne S.T. Rowe,~\IEEEmembership{Member, IEEE}}\\
\thanks{$^\star$Corresponding authors: D. Huang and A. Al-Hourani, E-mail: da.huang@ieee.org, akram.hourani@rmit.edu.au. Authors are with the School of Engineering, RMIT University, Melbourne, VIC 3000. This research received grant funding from the Australian Government under the Automotive Engineering Graduate Program. Copyright (c) 2024 IEEE. Personal use of this material is permitted. However, permission to use this material for any other purposes must be obtained from the IEEE by sending a request to pubs-permissions@ieee.org.}
}




\newpage

\begin{figure*}
{\huge Accepted by IEEE Internet of Things Journal}
\newline \newline
{\Large \textit{DOI: 10.1109/JIOT.2024.3361892}}
\newline \newline

© 20XX IEEE.  Personal use of this material is permitted.  Permission from IEEE must be obtained for all other uses, in any current or future media, including reprinting/republishing this material for advertising or promotional purposes, creating new collective works, for resale or redistribution to servers or lists, or reuse of any copyrighted component of this work in other works.
\end{figure*}

\newpage
\maketitle
\begin{abstract}
Transmitter authentication is critical for secured Internet of Things (IoT) applications. Recently, there has been growing interest in utilizing the physical layer authentication technique, radio frequency (RF) fingerprinting, to introduce extra security measurements without adding additional components. This work presents a novel fingerprint exploitation modality, Density Trace Plot (DTP), to leverage RF fingerprints originating from symbol transition trajectories for transmitter authentication. With a particular focus on IQ imbalance as the source impairment for RF fingerprints, we investigate the feasibility of three types of DTPs based on constellation, eye, and phase traces. The potential fingerprints presented in the DTP modalities are then used in training three deep learning classifiers: 2D-convolutional neural network (CNN), 2D-CNN+bi-directional long short-term memory (biLSTM), and 3D-CNN for transmitter authentication. The feasibility of the proposed approach in both wired and wireless conditions is validated using an experimental setup built using ADALM-PLUTO software-defined radios (SDRs). Experimental results demonstrate the best authentication accuracy of 96.7\% is achieved across signals of various modulation complexities. 
\end{abstract}


\section{Introduction}
Device authentication plays an important role in the security aspect of Internet of Things (IoT) systems. Popular end-device authentication techniques in IoT systems can be implemented either at the communication level, user level, or device level~\cite{Liyanage2020}. While the cryptography~\cite{Marin2015} and network flow~\cite{Zhao2023} oriented approaches serve as promising communication level approaches, they are found vulnerable to adversaries such as relay attacks~\cite{Francillon2011}. Similar vulnerability also applies to popular user level approaches that depend on biometrics~\cite{Reddy2020} or mutual authentication~\cite{liang2005performance}. In addition, these approaches usually require access to the application layer on both end devices, which might not be applicable to resource-constrained end-node devices.


In the meantime, device level approaches attempt to make use of low-level characteristics in assisting authentication. For instance, existing studies have explored introducing round-trip time (RTT)~\cite{Brelurut2016} 
and sensor fusion~\cite{Conti2020} to fingerprint devices. However, as high-end hardware might be in demand to efficiently put the above characteristics to use, recent literature has shown increasing interest in alternative physical layer authentication approaches like physical unclonable function (PUF) and radio frequency (RF) fingerprinting.

Approaches based on PUF require specific circuitry to be manufactured at the integrated circuit (IC) level~\cite{Chen2018PUF}, which can not be modified once manufactured. On the other hand, RF fingerprinting only utilizes effects due to imperfection factors, such as manufacturing error and device aging, to generate device-identifiable fingerprints. As such, it is more flexible and can be implemented across various systems without significant modifications to their existing architecture.
In RF fingerprinting, the device-identifiable fingerprints can be extracted from either the transient~\cite{Kose2019,Aghnaiya2020,Yang2022,Zhao2023a} or the steady state portion of the hardware-imperfection-affected received signals. Within steady state, typical wireless communication impairments such as phase noise (PN), carrier frequency offset (CFO),
IQ imbalance
and power amplifier (PA) imperfections~\cite{Mohammadian2021}
can be considered as potential sources of RF fingerprints.

This work explores the practicality of employing RF fingerprinting in authenticating end-node transmitters that utilize quadrature amplitude modulation (QAM) schemes. 
To achieve this, we propose density trace plot (DTP) as the signal representation modality for device fingerprint exploitation and introduce deep learning networks as the classifier. We further test the feasibility of the proposed DTP generated in both 2D and 3D formats and in three different domains. Furthermore, we consider the IQ imbalance as the only impairment source of device fingerprints. While relevant literature mainly considers IQ imbalance jointly with other impairments in forming mixture fingerprints, we explore the feasibility of leveraging IQ imbalance isolated from other impacts for transmitter authentication. This is because the impact of IQ imbalance on authentication performance is not addressed in previous literature. 
In addition, as the proposed framework only needs to be implemented on the receiver (e.g., base station) side, it provides an extra layer of security with minimum modification required for transmitters (i.e., end-node IoT devices). The main contribution of this work is summarized as follows:
\begin{itemize} 
    \item It proposes DTP as a fingerprint exploitation modality. In particular, the performance of 2D and 3D variations of constellation DTP, phase DTP and eye DTP are evaluated and compared.
    \item It validates the feasibility of utilizing both 2D and 3D deep learning algorithms as classifiers. Specifically, 2D-CNN, 2D-CNN+biLSTM and 3D-CNN are considered. 
    \item It validates the performance of the proposed framework under both wired and wireless practical setups. The experiments consider QAM signals with varying modulation orders. More specifically, QAM with modulation order ranging from 2 to 64 is studied.
\end{itemize}

The rest of the paper is structured as follows: Section~\ref{sec:BG} conducts a review of relevant literature. Section~\ref{sec:ME} presents the system framework and the proposed DTP method. Section~\ref{sec:ED} details the utilized practical setup and data collection procedure. Section~\ref{sec:RD} evaluates and discusses the performance of the proposed approach, and Section~\ref{sec:C} concludes the work.

\section{Related Work} 
\label{sec:BG}
Due to factors such as device aging or manufacturing errors, different devices embrace unique intrinsic impairments along their lifespan. As these intrinsic impairments interact in a device-independent stochastic manner, their resulting effects can be cultivated as device-identifiable yet hard-to-clone RF fingerprints. In literature, both a signal's transient and steady state segments can be used to exploit potential fingerprints. The transient-based approaches focus on the signal's short turn-on/off segment. This segment does not carry any actual data and only reflects the unique performance of the electronics. Features like energy spectral coefficients~\cite{Kose2019} and higher order statistical (HOS) measurements such as instantaneous profiles, skewness and kurtosis~\cite{Aghnaiya2020,Yang2022} can be calculated as potential RF fingerprints. However, a highly sensitive and well-synchronized receiver is required to acquire the rapid transients, making the implementation costly. The steady state fingerprints, however, are easier to obtain. Although HOS measurements~\cite{Wang2019b} remain a feasible option for steady state fingerprints, relevant studies tend to utilize specific impairments as the fingerprint sources. In a typical transceiver, the impairments that act as potential fingerprint sources can either be utilized separately~\cite{Hao2019,Hou2014,Li2021a} or considered together as a mixture of fingerprints~\cite{Pospisil2013,Sankhe2020,Peng2020}.


Despite primarily being a modulation domain impairment, IQ imbalance causes variations in various domains, making it a suitable source for RF fingerprints. In~\cite{Hao2019}, estimated IQ imbalance coefficients obtained through a generalized likelihood ratio test (GLRT) estimator were leveraged in a whitelist-based authentication approach for detecting transmitter relays. As IQ imbalance estimation modules have been typical components embedded in many off-the-shelf receivers, the authors highlight that utilizing IQ imbalance for authentication is cost-efficient. The study~\cite{Rondeau2021} focuses on authenticating Zigbee devices that transmit orthogonal quadrature phase shift keying (OQPSK) signals, where captured signals are projected into complex quadrant sequences that are further used to construct 72-dimension fingerprint vectors. The fingerprint vectors are subsequently used to train a multiple discriminant analysis and maximum likelihood (MDA/ML) model to achieve transmitter authentication. In other literature, the IQ imbalance is often considered jointly with other impairments to form combined fingerprints. For instance, in~\cite{Pospisil2013}, the received OQPSK signal is compared to its impairment-compensated version to generate an error vector that captures effects due to IQ imbalance, direct-current (DC) offset, CFO and PA non-linearity. A Gaussian mixture model (GMM) classifier is then employed to authenticate transmitters. In another work~\cite{Sankhe2020}, raw orthogonal frequency division multiplexing (OFDM) signals that encompass IQ imbalance, DC offset, carrier phase offset (CPO) and CFO are input into a two-channel 1D-CNN. This CNN outputs binary sequences as the coding of input signals. The coding sequences are then compared with the whitelist-based matching table for device classification.

A constellation domain presentation approach, namely differential constellation trace figure (DCTF), is proposed in a series of works~\cite{Jiang2019,Peng2020,Yang2021}. A new signal sequence is first obtained through a differential process between the original raw signal and its delayed and conjugated version. The DCTF is then constructed by projecting this signal onto a 2D constellation map. The generated DCTF mitigates CFO into a fixed parameter and considers DC offset, IQ imbalance and the stabilized CFO as its fingerprint sources.  The k-nearest neighbor (k-NN) and support vector machine (SVM) classifiers are introduced to learn features from the DCTF clustering centers~\cite{Jiang2019}. Alternatively, 2D-CNN is also adapted to handle 2D DCTF images directly~\cite{Peng2020}. In a similar work~\cite{Li2021}, the differential process is introduced to generate 2D colored contour images from WiFi signals, with 2D-CNN used as the classifier. In~\cite{Oligeri2023}, the IQ contour plot is generated from the 2D histogram to authenticate 4-QAM-like downlink signals of the satellites. However, the authors claim the approach requires a significant number of IQ samples per image for best performance. 
In the meantime, other signal transforms like short-time Fourier transform (STFT)~\cite{Huang2021, Shen2021a}, Gabor transform~\cite{Fadul2021}, continuous wavelet transform (CWT) and general linear chirplet transform~\cite{Baldini2020} both show their efficiency in exploiting cross-domain RF fingerprints. To assist in determining the most discriminating transform, efforts such as an ensemble method is proposed in~\cite {Baldini2023} to select the optimal transform based on their estimated Shannon entropy.

An efficient classifier is essential for accurate authentication. As previously reviewed, traditional learning techniques such as the MDA/ML~\cite{Rondeau2021,Wang2019b} and conventional machine learning techniques SVM~\cite{Aghnaiya2020,Wang2019b} and k-NN~\cite{Yang2021} are popular classifiers. However, as the physical layer impairments interact non-linearly, accurately extracting decisive RF fingerprints through manual feature engineering is challenging. Benefiting from the ability to extract features automatically with minimal manual interaction, deep learning classifiers have received growing research interest in recent literature. For instance, 1D-CNN~\cite{Sankhe2020,Jian2020,Fadul2021} and LSTM~\cite{Shen2021a} are commonly employed for 1D inputs like IQ sequences in transmitter authentication. In the meantime, 2D-CNNs are widely used to handle 2D inputs. Meanwhile, higher-dimension model 3D-CNN~\cite{Ramasubramanian2021} shows potential in learning temporally correlated fingerprints. It is also worth noticing that there are efforts to explore classifiers built using multi-stage architecture. For example, an autoencoder can be introduced in assisting initializing CNN's hyperparameter~\cite{Fadul2021}. In other approaches, CNN is used as the feature extractor, while another network such as CNN~\cite{Zhao2023a,Wu2023} or multilayer perceptron (MLP)~\cite{Zhang2023} is introduced to assist in fine-tuning the feature extractor and act as the actual classifier.


\section{Methodology}
\label{sec:ME}
The impact of physical layer impairments can be interpreted as tiny variations embedded in the received signal. Therefore, relevant processing like RF fingerprint exploitation and authentication needs to be implemented on the receiver side.
While various impairments could serve as potential sources of RF fingerprints, we primarily focus on intrinsic physical layer defection IQ imbalance in this work. We start by demonstrating that IQ imbalance's effects rely on the degree of hardware imperfection only rather than the effects caused by propagation. Thus, it can be considered a stable and location-irrelevant impairment. Accordingly, we generate DTPs from the pre-processed received signal for transmitter authentication. For modulation schemes with limited symbol patterns, DTP helps overlap the repetitive inter-symbol transition traces into the same representation space, such that the density within such space can be extracted as an additional fingerprint. Lastly, since DTP represents the signal as 2D or 3D representations, we adopt deep learning algorithms to serve as the classifiers. This is due to their excellent performance in automatic feature engineering when handling high-dimension inputs. 

\subsection{Overall System Model}
\begin{figure*}[htbp]
    \centering
    \includegraphics[width=0.6\linewidth]{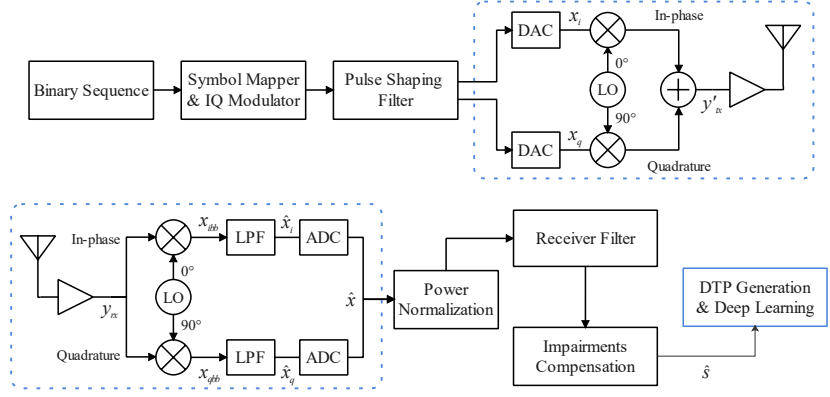}
    \caption{System model diagram of the proposed direct conversion single carrier system framework.}
    \label{fig:Sys}
\end{figure*}

In typical single-carrier direct-conversion transceivers, the local oscillator and its subsequent PAs generate a pair of orthogonal sine and cosine carrier waves. This orthogonal wave pair is used for up/down conversion of the in-phase (I) and quadrature (Q) channel signals. Theoretically, the orthogonal wave pair has identical phases and frequencies for both end devices. However, the presence of physical layer imperfections introduces impairments like CFO, CPO and IQ imbalance during signal transmission. Specifically, IQ imbalance arises as gain and phase mismatches between the orthogonal wave pair~\cite{Sankhe2020}. Consequently, IQ imbalance causes carrier waves of the I and Q channels to lose orthogonality. The overall system model diagram is shown in Fig.~\ref{fig:Sys}, indicating both the transmitter (device under test) and the receiver (detector device). Components within the dotted line boxes represent those embedded in the hardware. Since we show no interest in the actual payload content, no demodulating and decoding is needed within our system model.


At the transmitter end, the continuous baseband I and Q channel signal can be expressed as $x_\text{i}(t)$ and $x_\text{q}(t)$ respectively. Accordingly, an ideal bandpass signal out of the transmitter is given by
\begin{equation}
    y_\text{tx}(t)=\Re{\{x(t)e^{i2\pi f_\text{c}t}\}} = x_\text{i}(t)\text{cos}(2\pi f_\text{c}t)-x_\text{q}(t)\text{sin}(2\pi f_\text{c}t),
\end{equation}
where $f_\text{c}$ is the frequency of the carrier signal generated by the local oscillator. Due to physical layer imperfection, I and Q branch signals suffer from unequal carrier amplitude and phase, further known as the IQ imbalance. When taking the signal of the I channel as the reference, we denote the relative amplitude gain as $\alpha$ and the relative phase difference as $\phi$, which are associated with the Q signal. In such manners, an imbalanced RF analytical bandpass signal out of the transmitter is represented as
\begin{equation}
     y'_\text{tx}(t) = x_\text{i}(t)\text{cos}(2\pi f_\text{c}t)-\alpha x_\text{q}(t)\text{sin}(2\pi f_\text{c}t+\phi). \label{eq:IQibtx}  
\end{equation}



For an arbitrary channel with impulse response $h(t)$. The bandpass signal at the receiver can then be given as
\begin{equation} 
\begin{split}
    y_\text{rx}(t) & =h(t)*y'_\text{tx}(t)\\
     & = h(t)*x_\text{i}(t)\text{cos}(2\pi f_\text{c}t)-h(t)*\alpha x_\text{q}(t)\text{sin}(2\pi f_\text{c}t+\phi),
\end{split}
\end{equation}
where $*$ denotes the convolution operation. Without loss of generality, here we consider an ideal channel with a unity gain and zero phase shift (i.e., $h(t)=1$) for easy derivation.
\begin{figure*}[!tbhp]
\begin{align}
     x_\text{ibb}(t)&=y_\text{rx}(t)\text{cos}(2\pi f_\text{crx}t+\theta) \nonumber\\
     &=\frac{x_\text{i}(t)}{2}\left[\text{cos}(2\pi (2f_\text{crx}+\delta f)t+\theta)+\text{cos}(2\pi \delta ft-\theta)\right] \nonumber\\
     &-\frac{\alpha}{2}x_\text{q}(t)\left[\text{sin}(2\pi (2f_\text{crx}+\delta f)t+\phi+\theta)
     +\text{sin}(2\pi \delta ft+\phi-\theta)\right].
     \label{eq:rxi2}
\end{align}
\begin{align}
    x_\text{qbb}(t)&=y_\text{rx}(t)\beta\text{sin}(2\pi f_\text{crx}t+\psi+\theta) \nonumber\\
     &=\frac{\beta}{2}x_\text{i}(t)\left[-\text{sin}(2\pi (2f_\text{crx}+\delta f)t+\psi+\theta)
     +\text{sin}(2\pi \delta ft-\psi-\theta)\right] \nonumber\\    
     &+\frac{\alpha\beta}{2}x_\text{q}(t)\left[\text{cos}(2\pi (2f_\text{crx}+\delta f)t+\phi+\psi+\theta)-\text{cos}(2\pi \delta ft+\phi-\psi-\theta)\right].
     \label{eq:rxq2}
\end{align}
\begin{align}
      \hat{x}_\text{i}(t)=\frac{x_\text{i}(t)}{2}\text{cos}(2\pi \delta ft-\theta)-\frac{\alpha}{2}x_\text{q}(t)\text{sin}(2\pi \delta ft+\phi-\theta).
     \label{eq:rxi2_LPF}
\end{align}
\begin{align}
       \hat{x}_\text{q}(t)=\frac{\beta}{2}x_\text{i}(t)\text{sin}(2\pi \delta ft-\psi-\theta)-\frac{\alpha\beta}{2}x_\text{q}(t)\text{cos}(2\pi \delta ft+\phi-\psi-\theta).
     \label{eq:rxq2_LPF}
\end{align}
\end{figure*}
\noindent

In a typical receiver, the imperfections of the receiver introduce additional impairments during the down-conversion process, manifested as the CPO $\theta$, the CFO $\delta f=f_\text{ctx}-f_\text{crx}$, where $f_\text{ctx}$ and $f_\text{crx}$ respectively stand for the transmitter and receiver carrier frequency. In addition, receiver's IQ imbalance introduces amplitude mismatch $\beta$ and phase mismatch $\psi$. Consequently, the received baseband signals $x_\text{ibb}$ and $x_\text{qbb}$ are respectively given in (\ref{eq:rxi2}) and (\ref{eq:rxq2}) considering impairments as per mentioned above. The two signals are then filtered using low-pass filters (LPFs) to remove high frequency components, which further gives $\hat{x}_\text{i}$ and $\hat{x}_\text{q}$ in (\ref{eq:rxi2_LPF}) and (\ref{eq:rxq2_LPF}) respectively. Finally, after the analog-digital-converter (ADC), the discrete baseband analytical signal is given as $\hat{x}[n]=\hat{x}_\text{i}(nT_\text{s})-j\hat{x}_\text{q}(nT_\text{s})$, 

\subsection{Signal Pre-processing}
\label{sec:sigpro}
Given our primary interest in IQ imbalance, pre-processing is conducted to compensate other unwanted impairments and effects. Impairments to be compensated include CFO, CPO, DC offset and timing clock skewness. 

Due to transmission propagation, received signal $\hat{x}[n]$ undergoes power losses and is affected by channel noises. We remove the DC offset and normalize $\hat{x}[n]$ against its root mean square (RMS) to achieve unity power. We conduct energy detection to locate the presence of $\hat{x}[n]$ within the receiver frame by detecting rapid changes within the signal's power. To minimize inter-symbol interference (ISI) and limit transmission bandwidth, we employ typical square-root raised-cosine (SRRC) filter pair at both transmitter and receiver. 

Accordingly, the fast Fourier transform (FFT)-based technique~\cite{Wang2004} is used for coarse CFO compensation. The method searches for the frequency peak within the FFT magnitude of the signal's $\text{m}^{th}$ power as the coarse CFO.
Further, carrier synchronization block is conducted for fine CFO and CPO compensation. In this work, the fine compensation is implemented as a two-step process. Firstly, the signal is passed through a filter bank with impulse response given as
\begin{equation}
    h^{\Theta}_\text{MF}[n]=\overline{s_\text{ref}[N_\text{ref}-n+1]e^{j\Theta}},~\Theta\in[-\pi,\pi),
\end{equation}
where $\Theta$ is the additional phase shift, $s_\text{ref}[n]$ is a reference signal and $N_\text{ref}$ is the length of $s_\text{ref}$. More detail on $s_\text{ref}[n]$ can be found in Section~\ref{sec:TS}. 
Filters within the filter bank are constructed by introducing different $\Theta$s. After passing the signal through the filter bank, $\Theta$ that yields the largest absolute convolution output is selected as the estimated CPO. The intermediate signal $s_\text{int}[n]$ after this step is denoted as
\begin{equation}
   s_\text{int}[n]=\hat{x}[n]\text{exp}\left(-j~\underset{\Theta}{\mathrm{argmax}}\left|(s_\text{FS}*h^{\Theta}_\text{MF})[k]\right|\right).
\end{equation}
The precision of the estimation is largely affected by the selection of $\Theta$. Therefore, the second compensation step introduces a conjugate product estimator (CPE)~\cite{CHAUDHARI2018} to remove residual CPO and CFO. The CPE works by estimating the phase difference between the input signal and the conjugated version of $s_\text{ref}[n]$.

Lastly, we utilize a typical phase lock loop (PLL) built using a timing error detector (TED)~\cite{Mengali1997} to fix the timing clock skewness and subtract the mean value of the time domain signal from itself to remove DC offset. As such, the signal after the pre-processing is given as
\begin{multline}
    \hat{s}[n] = \frac{x_\text{i}(nT_\text{s})}{2}-\frac{\alpha}{2}x_\text{q}(nT_\text{s})\text{sin}(\phi)+\\
    j\left[\frac{\beta}{2}x_\text{i}(nT_\text{s})\text{sin}(\psi)+\frac{\alpha\beta}{2}x_\text{q}(nT_\text{s})\text{cos}(\phi-\psi)\right]
    \label{eq:Ssync}
\end{multline}
with whose phase can be obtained as 
\begin{equation}
    {\Phi}_{\hat{s}}[n]=\text{atan2}\left(\operatorname{Im}(\hat{s}[n]),\operatorname{Re}(\hat{s}[n])\right),
     \label{eq:rxphbb}
\end{equation}
where atan2 is the two-argument form of the arctan function, which allows $[0,2\pi)$ phase shift.

\subsection{Density Trace Plot}
The pre-processed $\hat{s}[n]$ is converted into \textit{trace plots} for better feature representation. Compared to a symbol scatter plot, the trace plot is better at visualizing transitions between adjacent symbols. As the trace plot is simply a remapping of the source signal, it inherits all the impairments presented in (\ref{eq:Ssync}). Here, we consider the following trace types:
\subsubsection{Type 1 (constellation trace)}
 The constellation trace is a common visualization technique that depicts symbol transitions within an IQ plane. The presence of IQ imbalance causes the locations of symbols to shift, resulting in a squeezed and tilted appearance of the constellation. 
\subsubsection{Type 2 (eye trace)}    
The traces of an eye diagram illustrate the temporal transition between consecutive symbols in a signal. When comparing the eye traces between I and Q channels, the IQ imbalance causes misalignment between several characteristics of the two eye shapes. For instance, the imbalanced amplitude results in varying eye heights and amplitudes, while the imbalanced phase leads to misalignment in jitter, eye-opening and rise/fall time.
 \subsubsection{Type 3 (phase trace)}      
As calculated in (\ref{eq:rxphbb}), the phase of a signal is also affected by IQ imbalance. In the constellation view, the phase mismatch causes symbols to offset from their ideal locations. Such effects in the overlapped phase plots are translated as shifts in the signal's concentrated phase states.

\begin{figure}[htbp]
    \centering
    \includegraphics[width=0.9\linewidth]{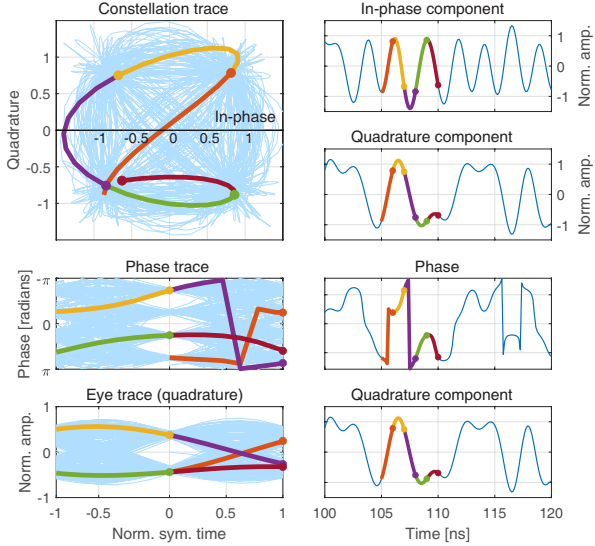}
    \caption{Example of utilized trace types. Light blue lines show the whole IQ trajectory of a 4-QAM sample, while the colored segments highlight the trajectories of five randomly selected consecutive symbols.}   
    \label{fig:traceExample}
\end{figure}

Example traces of an emulated 4-QAM signal are presented in Fig.~\ref{fig:traceExample}. The time domain waveform is normalized to have an amplitude between [-1,1] for both channels, and the phase and eye traces are plotted across a normalized time axis that covers two symbol durations. 


Nevertheless, the conventional trace plots become overcrowded as the length of the source signal increases and are less efficient in carrying information. To address this, we overlap symbol transition traces as density maps inside DTP, incorporating additional perspectives for the classifier to learn features from. 

During pulse shaping, the filter upsamples the input signal by samples-per-symbol ($sps$) times. Here, we reuse the upsampled signal for DTP generation to recover as many trajectories between adjacent symbols as possible. In this work, $sps=8$ is used as it suffices to generate adequate traces while avoiding significant computational overhead. To exploit density information, we map the discrete source sequence $s[n]$ into a bivariate histogram of width (number of columns) $w$ and height (number of rows) $h$. For Type 1 and 2 DTPs, $s[n]=\hat{s}[n]$, and $s[n]={\Phi}_{\hat{s}}[n]$ for Type 3 DTP. The bivariate histogram groups data into 2D bins and records the number of data points $v(r,c)$ that fall into a certain bin $b(r,c)$. That is, for arbitrary data point $(x_\text{p},y_\text{p})$, it falls into bin $b(r,c)$ if it satisfies both $x_\text{p}\in\left[x_\text{c},x_\text{c+1}\right)$ and $y_\text{p}\in\left[y_\text{r},y_\text{r+1}\right)$, where $r\in[1,h]$, $c\in[1,w]$, while $x_\text{c}$ and $y_\text{r}$ are the lower bounds of $b(r,c)$. More specifically, the lower bounds are determined by
\begin{equation}
   x_\text{c}=x_\text{min}+\frac{|x_\text{max}-x_\text{min}|(c-1)}{w}
\end{equation}
and
\begin{equation}
   y_\text{r}=y_\text{min}+\frac{|y_\text{max}-y_\text{min}|(r-1)}{h},
\end{equation}
where $[x_\text{min},x_\text{max}]$ and $[y_\text{min},y_\text{max}]$ are the theoretical axis boundaries presented in the traces plots of Fig.~\ref{fig:traceExample}. To summarize, the process of mapping a given $s[n]$ into the proper domain and generating the DTP is outlined in Algorithm~\ref{alg:dtp}.

\begin{algorithm}
    \caption{Generating DTP from $s[n]$}\label{alg:dtp}
    \begin{algorithmic}
    
        \State $v(r,c)=0~\text{for all bins}$
        \For{$idx \gets 1$ to length of $s[n]$}        
        \If{Type 1 DTP}
            \State $x_\text{p} \gets \operatorname{Re}(s[idx]),~y_\text{p} \gets \operatorname{Im}(s[idx])$
        \ElsIf{Type 2 or Type 3 DTP}
            \State $x_\text{p} \gets s[idx]$
            \State $y_\text{p} \gets (idx\mod2sps\times f_\text{s}/R_\text{sym})-sps\times f_\text{s}/R_\text{sym}$
        \EndIf 
        \For{all combinations of $(r,c)$}
        \If{$x_\text{c}\leq x_\text{p}<x_\text{c+1}$ and $y_\text{r}\leq y_\text{p}<y_\text{r+1}$}
            \State $v(r,c) \gets v(r,c)+1$
        \EndIf
        \EndFor
        
        \EndFor
    
    \end{algorithmic}
\end{algorithm}

To prevent extremely high bin values, $v(r,c)$ are log-normalized into a range of $[0,255]$. That is, for arbitrary $v(r,c)$, it is normalized to
\begin{equation}
    v(r,c)=\left\{\begin{array}{ll}
    \left\lceil\frac{255\times\text{log}_{10}(v(r,c))}{\text{log}_{10}(\text{max}(v(r,c)))}\right\rceil&,~\text{if}\ v(r,c)\neq0\\
    0&,~\text{if}\ v(r,c)=0
    \end{array}\right.,
    \label{eq:bin_norm}
\end{equation}
where $\text{max}(\cdot)$ denotes finding the global maximum.

\begin{figure}[htbp]
    \centering
    \includegraphics[width=0.9\linewidth]{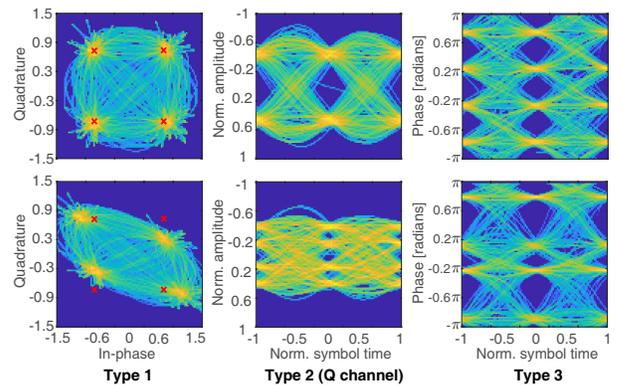}
    \caption{Examples 2D-DTPs of a 4-QAM sample, with SNR = 10~dB. The upper row presents DTPs generated from the original sample, and the bottom row presents DTPs generated when injected with emulated IQ imbalance.
    }
    \label{fig:DTPs}
\end{figure}

\begin{figure}[htbp]
    \centering
     \includegraphics[width=0.8\linewidth]{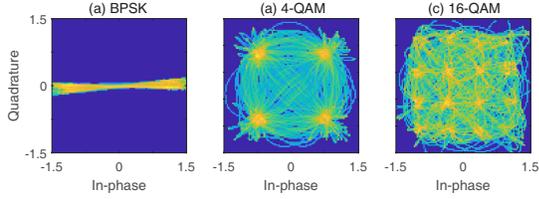}
    \caption{Example Type 1 DTPs generated for different modulation orders. All are generated based on 500-symbol-long sequences with SNR = 10~dB.}
    \label{fig:constDTPs}
\end{figure}

Examples of how DTPs react to IQ imbalance and varying modulation orders are presented in Fig.~\ref{fig:DTPs} and Fig.~\ref{fig:constDTPs}, respectively.
All examples are presented using pseudo-colors only for better visualization, while the actual DTP is generated as a single-channel 2D matrix. As a special case, the eye (Type 2) DTP for I and Q channels are concatenated after being generated separately. Consequently, the dimension of Type 2 DTP is $[h \times w \times 2]$, while remains $[h \times w \times 1]$ for constellation (Type 1) and phase (Type 3) DTPs.

\begin{figure}[htbp]
    \centering
\includegraphics[width=0.7\linewidth]{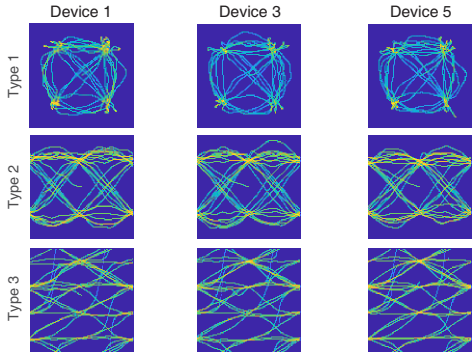}
    \caption{Example of variations across different transmitter devices. Generated from 4-QAM samples with 50 symbols at SNR of 10~dB. Note only the modality for Q-channel data is presented for Type 2 DTP.}
    \label{fig:transition}
\end{figure}

In addition, an example is given in Fig.~\ref{fig:transition} to showcase how different transmitters affect DTP. For clearer presentation, DTPs in this example are generated from the first 50 symbols of an identical 4-QAM signal propagated and captured through the same medium. It can be observed that the variations (i.e., fingerprints) across devices mainly manifest as the variation of certain density centers and ``outliers" transition traces.

\begin{figure}[htbp]
    \centering
    \includegraphics[width=0.85\linewidth]{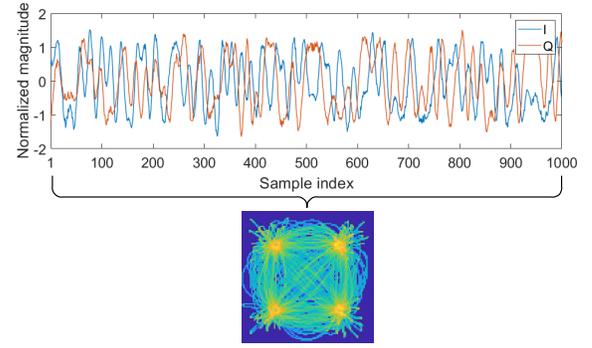}
    \caption{Example of Type 1 2D-DTP generated from a 4-QAM sample, with SNR = 10~dB.}
    \label{fig:2DDTP}
\end{figure}

\begin{figure}[htbp]
    \centering
\includegraphics[width=0.85\linewidth]{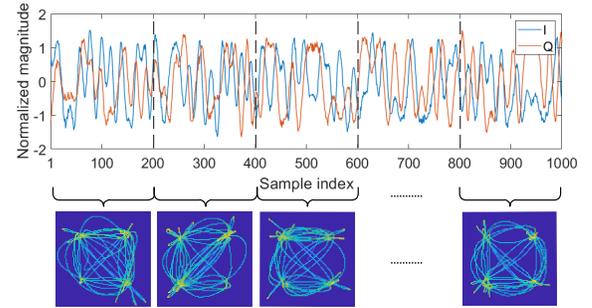}
    \caption{Example of constituting Type 1 3D-DTP using a sequence of 2D-DTP frames from a 4-QAM sample, with SNR = 10~dB.}
    \label{fig:3DDTP}
\end{figure}

The examples presented in Fig.~\ref{fig:DTPs} and Fig.~\ref{fig:constDTPs} demonstrate generating a DTP using the entire signal. This method is further referred to as 2D-DTP, as illustrated in Fig.~\ref{fig:2DDTP}. In this approach, all temporal variations are projected onto spatial 2D representations, akin to capturing a moving object using a long-exposure camera shot. Alternatively, the same signal can be divided into $N_\text{f}$ frames (segments), with each segment generating a 2D-DTP. All the generated 2D-DTPs are stacked to create a 3D sequence, as illustrated in Fig.~\ref{fig:3DDTP}. This method is further named after 3D-DTP, which is proposed exclusively for 3D deep learning models as an attempt to extract additional temporal fingerprints from the same source signal. Consequently, the dimensions of the 3D-DTP become $[h\times w \times N_\text{f} \times 1]$ for Type 1 and Type 3 DTP, while they are $[h\times w \times N_\text{f} \times 2]$ for Type 2 DTP. It is obvious that the selection of $h$, $w$ and $N_\text{f}$ impacts the resolution of both 2D- and 3D-DTP. The impact of these parameters is further discussed in Section~\ref{sec:parms}.

\subsection{Classifier}
We introduce three different deep learning networks in this work to handle DTPs of various dimensions and types. All proposed network architectures are built based on fundamental function blocks, whose number $n_\text{b}$ can be dynamically adjusted to meet different design requirements.
\subsubsection{2D-CNN}
2D-CNN is a feed-forward deep algorithm and an efficient tool for 2D input classifications. We design a multi-stage 2D-CNN similar to the one proposed in our previous work~\cite{Huang2021} to handle 2D-DTPs. As shown in Fig.~\ref{fig:Nwks}(a), each function block contains two 2D convolutional layers activated by ReLU and batch normalization layers and followed by a $2\times2$ max-pooling layer to reduce feature map dimensions.

\subsubsection{2D-CNN+biLSTM}
Since 2D-CNN only handles 2D inputs, additional networks are implemented to make the most of 3D-DTPs. Inspired by~\cite{Skaria2020}, we concatenate a bidirectional long short-term memory (biLSTM) block to the 2D-CNN to generate a hybrid architecture. The biSLTM is a variation of LSTM that learns features in both forward and backward directions. It is more efficient than the conventional LSTM network if the complete sequence is always available. As the 3D-DTP sequences are always generated using the entire $\hat{s}[n]$, they are hence suitable for biLSTM. As shown in Fig.~\ref{fig:Nwks}(b), the CNN network sequentially processes each frame within the 3D-DTP. It flattens and stacks all learned activations after the last max-pooling layer and produces an intermediate feature matrix with dimensions of $[N_\text{ac}\times N_\text{f}]$, where $N_\text{ac}$ represents the number of activations from the last max-pooling layer. This intermediate feature matrix is treated as a time series block with $N_\text{f}$ time points and $N_\text{ac}$ learnable features per time point and is used as input for the biLSTM network.

\subsubsection{3D-CNN}
\begin{figure}[htbp]
    \centering
    \includegraphics[width=0.75\linewidth]{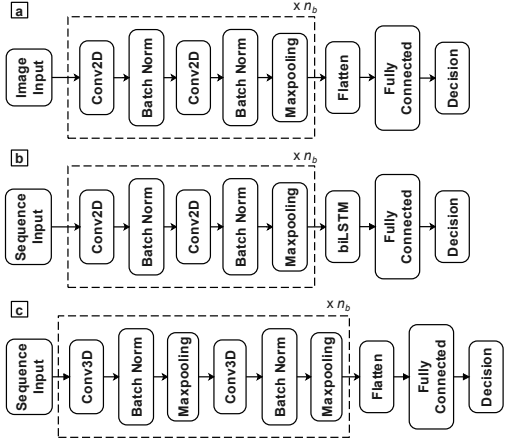}
    \caption{Example architectures of three utilized deep learning networks: (a) 2D-CNN, (b) 2D-CNN+biLSTM and (c) 3D-CNN.}
    \label{fig:Nwks}
\end{figure}

As an alternative, the 3D-CNN is utilized to handle 3D-DTP sequences. As in Fig.~\ref{fig:Nwks}(c), the architecture of a 3D-CNN is similar to the 2D-CNN but with 3D convolutional layers replacing the 2D ones. Unlike 2D-CNN+biLSTM, the 3D-CNN takes the entire 3D-DTP sequence at once. Due to this nature, the training of 3D-CNN is memory-demanding. To prevent potential memory depletion, the max-pooling layers are placed after every batch normalization layer to reduce the feature map's size rapidly.


\section{Experimental Setup and Data Handling}
\label{sec:ED}
\subsection{Practical Hardware Setup}

\begin{figure}[htbp]
    \centering
    \subfloat[]{
    \includegraphics[width=0.48\linewidth]{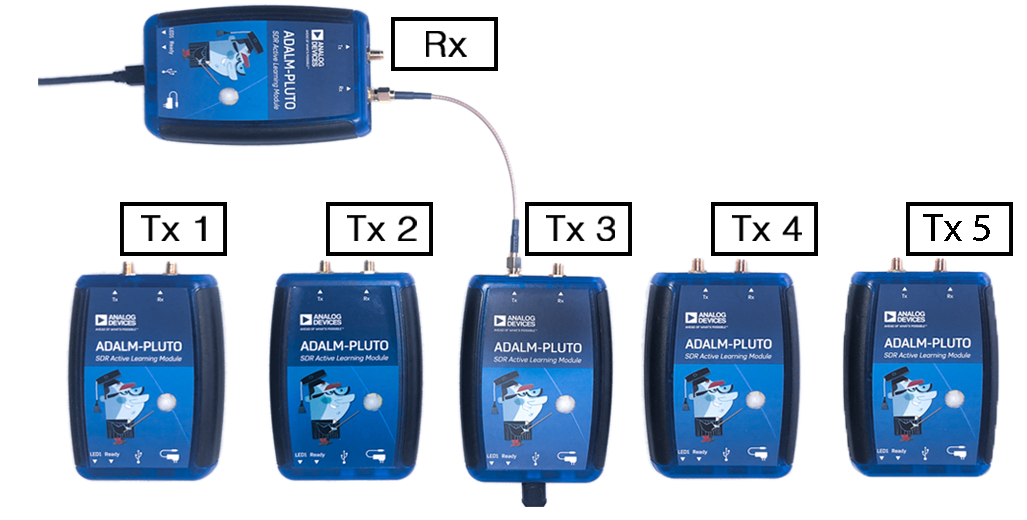}
    }    
    \subfloat[]{
    \includegraphics[width=0.48\linewidth]{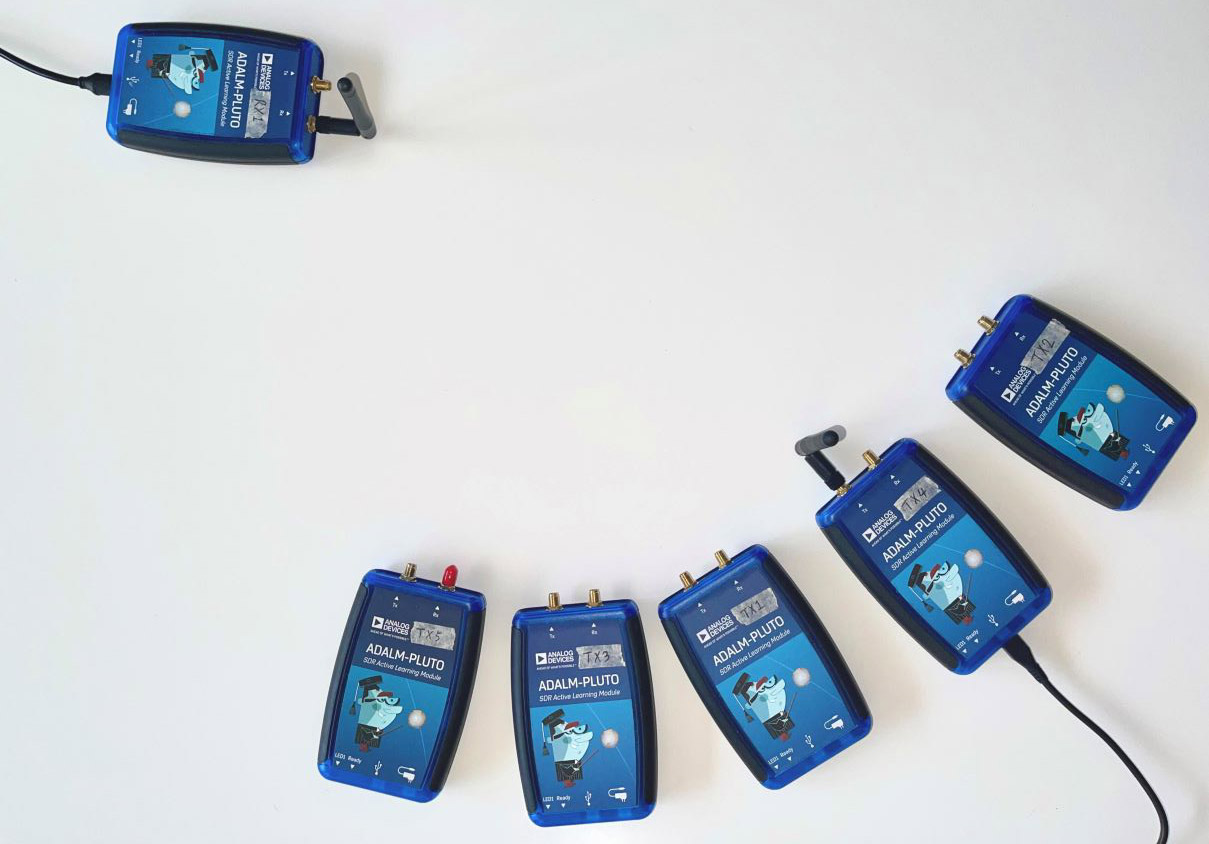}
    }
    \caption{Experimental hardware setup contains the utilized transmitters and receiver. Examples of (a) a wired connection and (b) an over-the-air channel.}
    \label{fig:HW}
\end{figure}

Our experimental setup uses the commercially available ADALM-PLUTO software-defined radio (SDR) as both transmitter and receiver devices. We select this hardware, rather than other off-the-shelf IoT modules, as it grants us more configuration and processing flexibility. 
Five SDRs are consistently used as transmitters, plus an additional SDR acts as the receiver. As in Fig.~\ref{fig:HW}(a), the wired setup is established by directly connecting the transmitter and receiver SDRs using an RF cable. This configuration minimizes wireless channel effects, allowing us to focus primarily on effects due to physical layer imperfections. Subsequently, the over-the-air test is conducted to validate the proposed approach in a more practical setup. Similar to Fig.~\ref{fig:HW}(b), the transmitters and the receiver are positioned inside the same room at an approximate distance of three meters. Given the presence of a line of sight (LoS) path, the channel involved in this setup can be regarded as a Rician channel.

The transmitter and receiver are connected to the same PC but controlled by separate MATLAB instances to allow parallel operation. The auto gain controller (AGC) is disabled on devices of both ends. 
Additionally, noticed in~\cite{ALHOURANI2018317}, potential flicker noise occurs around DC (i.e., 0Hz) on the receiver end. Thus, to avoid operating near the center of the baseband spectrum, a frequency shift $f_\text{cs}$ is applied to the transmitter so that gives $f_\text{ctx}=f_\text{c}+f_\text{cs}$. The $f_\text{cs}$ is removed at the receiver during pre-processing, yielding a baseband signal whose spectrum is located at a distance from the flicker noise.

In addition, we make a few assumptions and measurements to make the experimental setup more controllable. As pointed out in~\cite{Filler1993}, significant oscillator degradation (e.g., frequency drift exceeding 10ppb) might take around a year to manifest. Based on this, we assume the impairments remain relatively constant for all tested devices across the experiment period. Under this assumption, $\beta$ and $\psi$ due to receiver IQ imbalance can be regarded as fixed constants in (\ref{eq:Ssync}), enabling us to focus only on relative effects from the transmitters. Considering local oscillator behavior being sensitive to ambient temperature~\cite{Pospisil2017}, all data collection is conducted in an air-conditioned environment to prevent potential noise due to temperature fluctuations.

\subsection{Dataset Generation} \label{sec:TS}
We test signals modulated using binary phase shift keying (BPSK) (equivalent to 2-QAM), 4-QAM, 16-QAM and 64-QAM at a symbol rate of $R_\text{sym}=1$~Msymbol/s. Note the 64-QAM is only included during the over-the-air testing environment. The transmitted signals are generated as 500-symbol-long for all modulation schemes. This length adequately expresses sufficient fingerprints and only takes 0.5ms to transmit.
The signals for each modulation scheme are pseudo-randomly generated. After one-off generation, the identical signals are consistently reused throughout the test, maintaining consistent fingerprint expression during all transmissions. The ideal pulse-shaped signals are stored as the reference signals $s_\text{ref}$ and are used during pre-processing. 

The wired dataset is collected across five different days, where samples collected from the first four days are used for training, and samples collected on the fifth day are only used to test trained networks. Approximately 2,000 samples per modulation type per transmitter are collected during the first four days. On the fifth day, 500 samples are collected for each transmitter and each modulation scheme. The over-the-air dataset is collected in the same manner across another five days. In total, approximately 2,500 samples per modulation per transmitter are collected wirelessly. As a result, we collected a dataset comprising around 87,000 samples in total. 

As a consequence of propagation, transmissions reach the receiver with varying powers. To better quantify the relative impact of noise power, controlled additive white Gaussian noises (AWGNs) are injected into RMS normalized signal (prior to DTP generation) to achieve signal-to-noise (SNR) levels between -20~dB to 20~dB. This process allows us to produce a larger dataset with signals of different qualities.



\subsection{Training}
A random 80\% of the dataset is selected as \textit{training} set, and the remaining 20\% is set aside as the \textit{testing} set. In the \textit{training} set, the data is divided into three splits and are used for k-fold verification (i.e., one split performs as the \textit{validation} set each time). The training set only contains samples with SNRs of -10~dB, 0~dB and 10~dB, while samples in the testing set cover the full SNR range. 

As different DTP types embody fingerprints in different manners, deep learning algorithms are separately trained and optimized for each possible combination. During training, the Bayesian optimization technique is used to optimize the hyperparameters. The particular hyperparameters to be tuned include the number of function blocks $n_\text{b}$, the kernel size and filter numbers for each convolutional layer, and the number of hidden units for the biLSTM block. All the training is conducted using MATLAB on the PC with an Intel Core i7 10700KF CPU and one NVIDIA RTX GeForce 2070s GPU. During the training, the stochastic gradient descent with momentum (SGDM) is utilized as the solver. During optimization, a total of 200 iterations of hyperparameter sweep are conducted, with a maximum of 40 epochs per iteration. The trained network with the highest validation accuracy is selected as the final product for each possible combination.

Overall, we find the networks achieved a balance between performance and computational complexity when having $n_\text{b}=2$. Due to the large number of networks trained, presenting all their hyperparameters is impractical. Instead, we summarize the major hyperparameters of each modulation scheme's best-performed network in Table~\ref{hyperT}. Please refer to Section~\ref{sec:RD} for more details on comparing different networks.

\subsection{Parameters} \label{sec:parms}
\begin{figure}[htbp]
    \centering
    \includegraphics[width=0.7\linewidth]{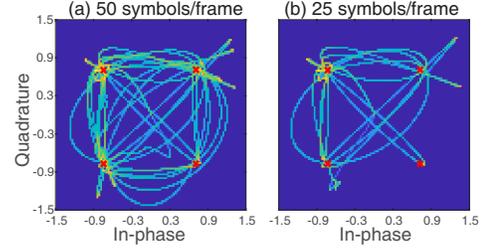}
    \caption{Example of Type 1 3D-DTP frames of a 4-QAM sample generated with different symbols-per-frame: (a) 50 symbols/ frame, and (b) 25 symbols/ frame. All examples are with $h=w=100$. The red crosses indicate the ideal position of each symbol.}
    \label{fig:DTPf}
\end{figure}

\begin{figure}[htbp]
    \centering
    \includegraphics[width=0.45\linewidth]{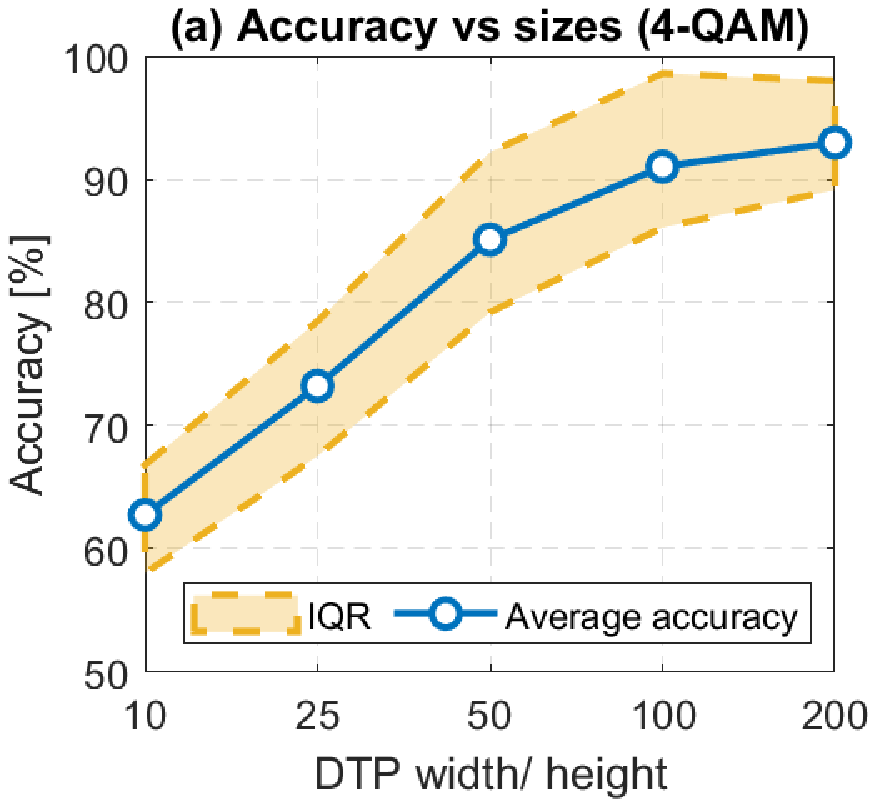}
    \includegraphics[width=0.45\linewidth]{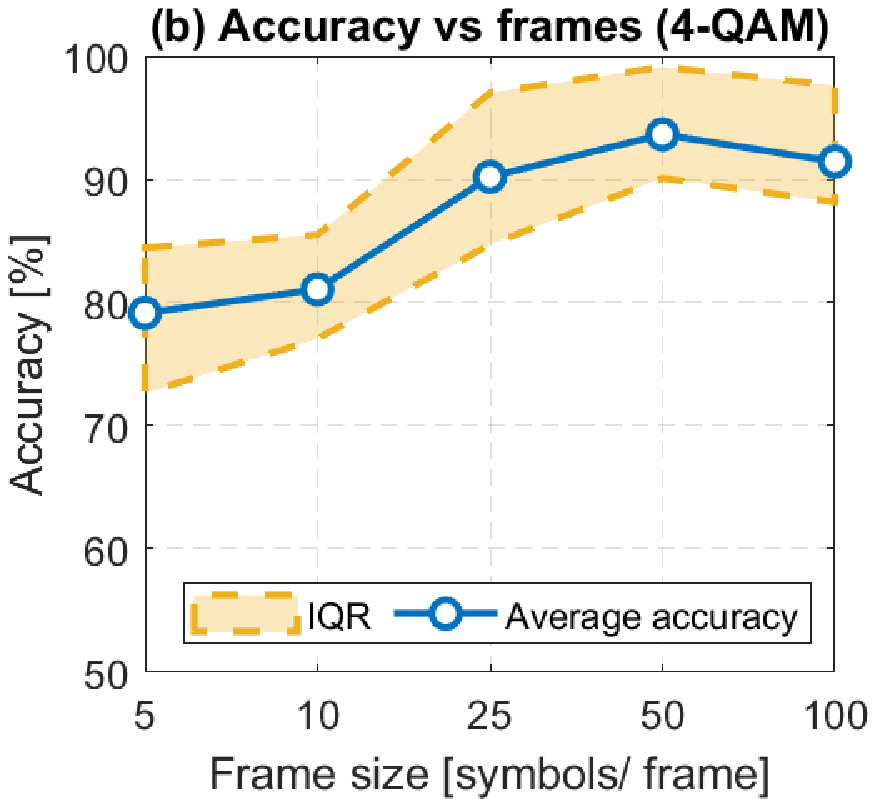}
    \caption{Preliminary results obtained at different parameter settings (SNR=10~dB), where (a) 2D-CNN trained using 2D-DTP of different $[h,w]$, and (b) 2D-CNN+biLSTM trained using 3D-DTP of different frames per symbols, when $h=w=\text{100}$.}
    \label{fig:DTPhwf}
\end{figure}

As mentioned previously, the selection of $[h,w]$ and $N_\text{f}$ affect the DTP's resolution and its capability to exploit fingerprints. The selection of $[h,w]$ impacts the number of bins available within the DTP modality. while in 3D-DTP, different frame numbers $N_\text{f}$ lead to varying symbols per frame. As illustrated in Fig.~\ref{fig:DTPf}, each frame manifests more traces with higher symbols/frame, i.e., smaller $N_\text{f}$. 
In order to get more insight into the impact, a subset of 4-QAM samples from all five transmitter SDRs is selected for a preliminary trial. For simplicity, only Type 1 DTP samples at SNR=10~dB are used. Following the training manner described in the previous section, 2D-CNN and 2D-CNN+biLSTM networks of different parameter settings are separately trained and tested. The evaluated results are presented as the interquartile range (IQR) and the mean classification accuracy as in Fig.~\ref{fig:DTPhwf}.

Based on the observation in Fig.~\ref{fig:DTPhwf}(a), we select $h=w=100$ for a balance between performance and complexity, given no significant accuracy increment is observed beyond this point. Additionally, we select 50 symbols/frame as it returns the highest accuracy from results presented in Fig.~\ref{fig:DTPhwf}(b). Since all source signals have an identical length, this selection corresponds to $N_\text{f}=10$. 

Values of other key parameters are summarized in Table~\ref{paramsT}, while the hyperparameters of each modulation scheme's best-performed network are summarized in Table~\ref{hyperT}. Every entry of Table~\ref{hyperT} follows the following format: \textit{[kernel width $\times$ kernel height] - filter numbers}. The approximate total learnable of each network is also provided, considering five classes.
\begin{table}[!htbp]
	\caption{Parameters table}
	\centering
	\begin{tabularx}{3.49in}{l X X}
		\hline\hline \\[-2ex]
		Parameter         				 &Symbol 			&Quantity\\
		\hline\\ [-2ex]
        Carrier frequency     & $f_\text{c}$     & 2.5 [GHz]  \\[-0.25ex]
        Carrier frequency shift & $f_\text{cs}$  & 1 [MHz]   \\[-0.25ex]
        Sampling frequency    & $f_\text{s}$     & 8 [MHz]    \\[-0.25ex]
        Symbol rate           & $R_\text{sym}$   & 1 [Msyms/s] \\[-0.25ex]
        Samples per symbol$^{1}$ & $sps$          & 8        \\[-0.25ex]
        DTP width (column number)            & w                & 100     \\[-0.25ex]
        DTP height (row number)           & h                & 100   \\[-0.25ex]
        Number of frames$^{2}$       & $N_\text{f}$     & 10 \\[-0.25ex]
        Number of convolutional blocks & $n_\text{b}$ & 2 \\[-0.25ex]
        Length of transmitted signal & $N_\text{s}$ & 500 \\[-0.25ex]
        Length of reference signal$^{3}$ & $N_\text{ref}$   & 500\\[-0.25ex]
		\hline 
	\end{tabularx}
\begin{tabularx}{3.49in}{l} \\[-1.5ex]
    $^1$ For pulse shaping filter.\\
    $^2$ For 3D-DTP generation, when having 50 symbols-per-frame.\\
\end{tabularx}
	\label{paramsT}
\end{table}


\begin{table}[htbp]

\caption{Hyperparameters for best-performed CNNs}
\centering
\begin{tabular}{ccccc}
\hline\hline \\[-2.5ex]
\multirow{2}{*}{Layers}                                            & \multicolumn{4}{c}{Hyperparameters} \\ \cline{2-5} \\ [-2ex]
                                                                   & BPSK    & 4QAM    & 16QAM  & 64QAM  \\ \hline\\ [-2ex]
conv\_1\_1                                                         & [8$\times$8]-22  & [5$\times$5]-25  & [7$\times$7]-24 & [7$\times$7]-24 \\
conv\_1\_2                                                         & [8$\times$8]-22  & [5$\times$5]-25  & [7$\times$7]-24 & [7$\times$]-24 \\
maxpool\_1                                                         & [2$\times$2]     & [2$\times$2]     & [2$\times$2]   & [2$\times$2]    \\
conv\_2\_1                                                         & [4$\times$4]-11  & [4$\times$]-7   & [4$\times$4]-27 & [7$\times$7]-23 \\
conv\_2\_2                                                         & [4$\times$4]-11  & [4$\times$4]-7   & [4$\times$4]-27 & [7$\times$7]-23 \\
maxpool\_2                                                         & [2$\times$2]     & [2$\times$2]     & [2$\times$2]    & [2$\times$2]    \\ \hline\\ [-2ex]
\begin{tabular}[c]{@{}c@{}}Approx.\\ total leanrables\end{tabular} & 41k     & 21.7k   & 58.3k  & 88.2k  \\ \hline
\end{tabular}
\label{hyperT}
\end{table}

\section{Results and Discussions}
\label{sec:RD}
\subsection{Behavior under Wired Transmission Link}


\begin{figure*}[!tbhp]
    \centering
    \includegraphics[width=0.75\linewidth]{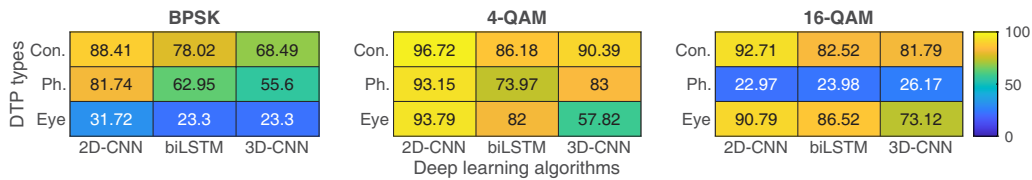}
    \caption{Classification accuracy (\%) of the constellation (Con.), phase (Ph.) and eye DTPs against different deep learning algorithms at SNR = 10~dB.}
    \label{fig:DTPaccs}
\end{figure*}

\begin{figure*}[!tbhp]
    \centering
    \includegraphics[width=\linewidth]{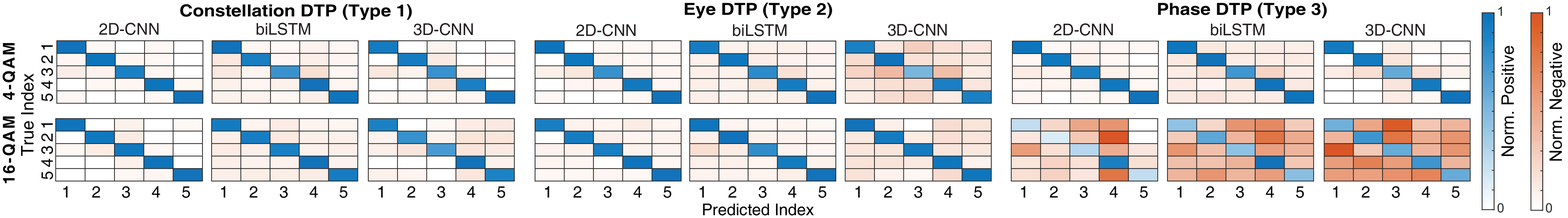}
    \caption{Confusion matrices for different DTP and deep learning pairs at SNR = 10~dB. Matrices are color-coded to showcase normalized positive rate and negative rate in different colors.}
    \label{fig:CMQAM}
\end{figure*}

\begin{figure*}[!tbhp]
    \centering
    \includegraphics[width=\linewidth]{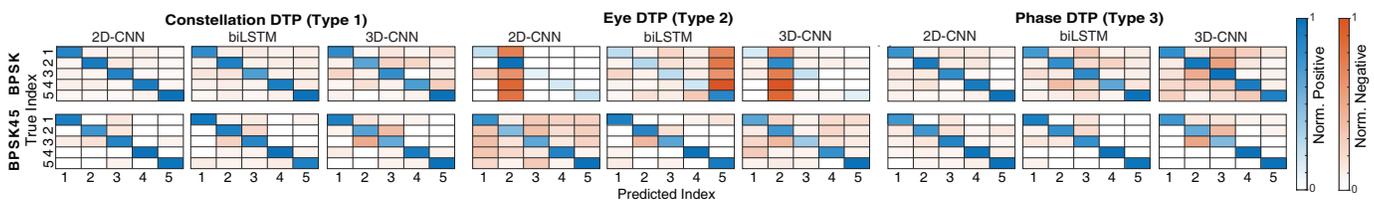}
    \caption{Confusion matrices for different DTP and deep learning pairs at SNR = 10~dB. Matrices are color-coded to showcase normalized positive rate and negative rate in different colors. The BPSK45 stands for phase-shifted BPSK.}
    \label{fig:CMBPSK}
\end{figure*}
Average classification accuracy of trained algorithms validated at SNR of 10~dB is summarized as heatmaps in Fig.~\ref{fig:DTPaccs}. Note as five transmitter devices are involved in the testing, the worst-scenario classification accuracy is $20\%$ (i.e., 1/5). Among all tested pairs, the Type 1 DTP yields the best overall average accuracy for all tested modulation schemes, especially the 2D-DTP and 2D-CNN pair. While the other two DTP types perform well in many cases, significant classification failures happen for the pair between Type 2 DTP with BPSK, as well as Type 3 DTP with 16-QAM.

To gain more insight into the device-wise classification performance, confusion matrices measured at SNR=10~dB are presented in Fig.~\ref{fig:CMQAM} and Fig.~\ref{fig:CMBPSK}. The matrices are row-normalized (i.e., respectively against the total observations of the corresponding true class) to visualize the \textit{recall} of a specific class. It is noted that misclassifications occur more frequently for certain devices, such as transmitter No.3. The varying performance between devices under test indicates that fingerprints of specific devices are less distinct and challenging for the trained networks to exploit.

\begin{figure}[tbhp]
    \centering
    \includegraphics[width=\linewidth]{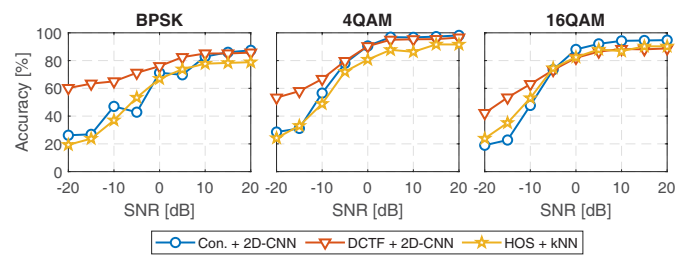}
    \caption{Performance comparison of the best-performed pairs for each modulation scheme under the wired channel.}
    \label{fig:topACC}
\end{figure}

An SNR sweep is further applied to the best-performing DTP and deep learning classifier pair of each modulation scheme. It is to analyze the proposed method's performance under various signal qualities. In addition, we compare the performance of DTP against DCTF~\cite{Peng2020} and HOS~\cite{Yang2022} representations generated from the same dataset. Since DCTF is also a 2D representation method, the 2D-CNN networks summarized in Table~\ref{hyperT} can be directly utilized. The HOS, on the other hand, produces twelve numerical features for each input, and hence kNN is used as the corresponding classifier. In the results shown in Fig.~\ref{fig:topACC}, a proportional relationship is found between the authentication accuracy and the SNR for both methods until the saturation happens at around SNR=10~dB. Under the same testing conditions, it is observed that although the DCTF achieves higher accuracy in low SNR$\leq$-10~dB, our proposed methods demonstrate similar or superior performance in high SNR regions, especially for higher modulation order schemes like 16-QAM. While HOS demonstrates a less accurate authentication performance in general.


\begin{figure}[!htbp]
    \centering
    \includegraphics[width=0.65\linewidth]{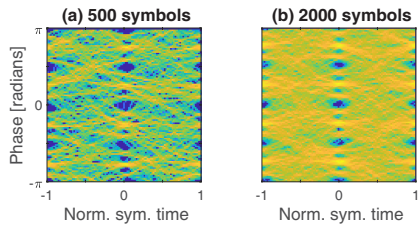}
    \caption{16-QAM Type 3 2D-DTP with SNR=10~dB generated when: (a) input signal is 500 symbols long, and (b) input signal is 2000 symbols long.}
    \label{fig:16QAMP}
\end{figure}

\begin{figure}[!htbp]
    \centering
    \includegraphics[width=\linewidth]{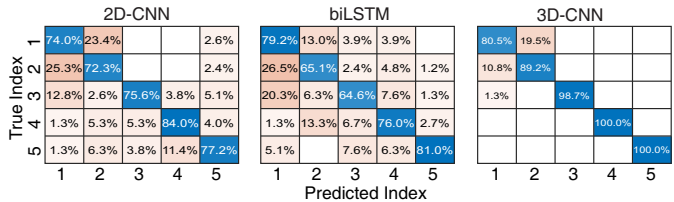}
    \caption{Confusion matrices for 16-QAM Type 3 DTPs with SNR=10~dB generated when the source signal is 2000 symbols long.}
    \label{fig:CM_16QAM2ms}
\end{figure}

As previously noted, a failure of Type 3 DTP is observed for 16-QAM. It may be due to the 16-QAM signals having more phase states and producing distinct density centers for a relatively short source signal is challenging. This ambiguity might result in inadequate expression of fingerprints. To test this assumption, a 16-QAM signal comprising 2000 symbols is generated to increase each symbol's occurrence. As shown in Fig.~\ref{fig:16QAMP} (b), the newly generated DTP shows more concentrated density centers along each phase state. A new set of networks is trained and tested using DTPs generated from the longer signal, and the corresponding results at SNR of 10~dB are presented in Fig.~\ref{fig:CM_16QAM2ms}. Compared to previous results, the classification accuracy of the longer signal is significantly improved, especially when paired with 3D-CNN.

\begin{figure}[htbp]
    \centering
    \includegraphics[width=0.85\linewidth]{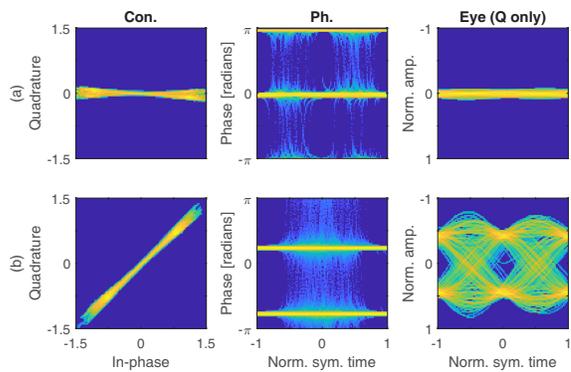}
    \caption{Example 2D-DTPs of BPSK sample with SNR=10~dB: (a) without phase shift, and (b) with 45$ ^{\circ}$ phase shift.}
    \label{fig:BPSK45DTPs}
\end{figure}

Similarly, significant failure is observed in Type 2 DTPs for BPSK, while the Type 1 and Type 3 DTPs also achieved worse overall performance than other modulation schemes. One possible reason is that ideal BPSK signals modulate no information on the Q channel, leading to $x_\text{q}(t)=0$ in (\ref{eq:rxi2_LPF}) and (\ref{eq:rxq2_LPF}), which consequentially reducing the amount of learnable fingerprints. To address this limitation, we attempted to introduce an additional 45-degree phase shift into BPSK signals before transmission. According to the examples in Fig.~\ref{fig:BPSK45DTPs}, the phase-shifted BPSK starts to employ more notable transitions, specifically for Type 2 DTPs.

\begin{figure}[htbp]
    \centering
    \includegraphics[width=0.65\linewidth]{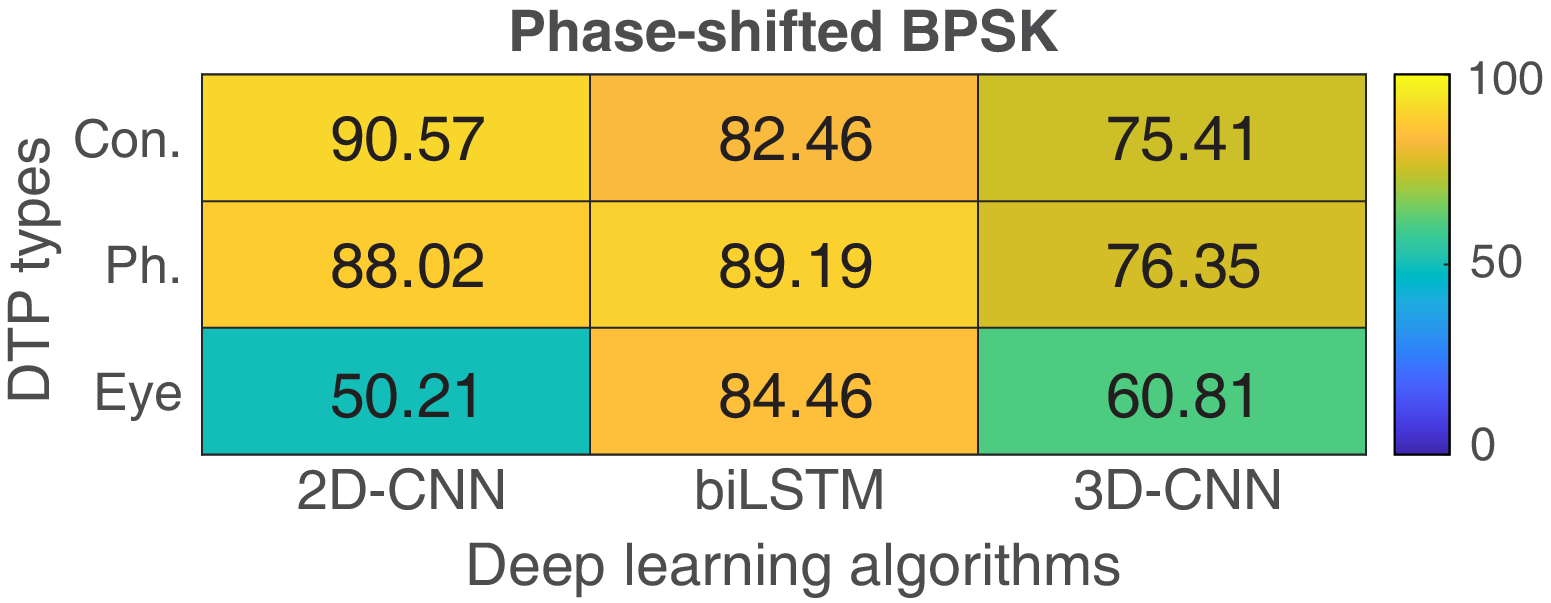}
    \caption{Classification accuracy (\%) of phase-shifted BPSK at SNR=10~dB.}
    \label{fig:BPSK45ACC}
\end{figure}

The phase-shifted BPSK samples are then used for another preliminary trial. After training another set of networks in the same manner, the testing results are respectively presented in Fig.~\ref{fig:CMBPSK} and Fig.~\ref{fig:BPSK45ACC}. More specifically, Fig.~\ref{fig:BPSK45ACC} presents the average authentication accuracy for all five transmitter devices at SNR of 10~dB, and the second row of Fig.~\ref{fig:CMBPSK} presents the confusion matrices for device-wise behavior evaluation. While the overall performance of the phase-shifted BPSK remains relatively worse compared to other modulation schemes, notable improvements can be observed if compared to the original BPSK modulation.

\subsection{Behavior under Wireless Transmission Link}


\begin{figure}[htbp]
    \centering
    \includegraphics[width=0.9\linewidth]{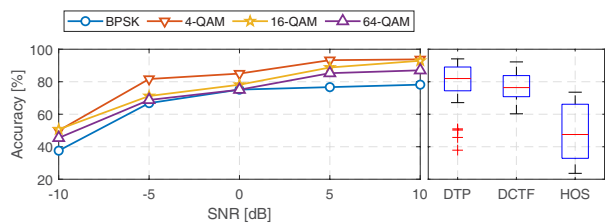}
    \caption{Performance comparison of the best-performed pairs for each modulation scheme under the presence of a Rician channel. Note both modulation adopts Type 1 DTP with 2D-CNN network.}
    \label{fig:topACCwl}
\end{figure}


Under the wireless scenario, we extended our testing to include 64-QAM samples. 
By utilizing the network architectures in Table~\ref{hyperT} and following similar training procedures, new 2D-CNN networks for wireless samples are trained and tested. 

The SNR sweep results for the wireless scenario are presented in Fig.~\ref{fig:topACCwl}. A similar comparison is again made between DTP, DCTF and HOS under the wireless condition. For better visualization, we plot the method comparison separately as box plots alongside the DTP SNR sweep. Both box plots are generated based on results from all modulation orders and SNR levels tested in this work.
The obtained results demonstrate that the wireless scenario yields similar results to the wired scenario despite slight drops in overall authentication accuracy due to a stochastic channel. Although the DCTF remains more robust within low SNR regions, the overall authentication accuracy of our proposed approach starts to outperform DCTF for SNR$\geq$-5~dB, particularly for high-order scheme 64-QAM. One possible explanation for the consistent performance of DCTF in low signal power regions is that the differential process during its generation tends to smooth out the impact of rapid variations caused by burst noise and contributes to a more stable representation.

\begin{figure}[!htbp]
\centering
\includegraphics[width=0.95\linewidth]{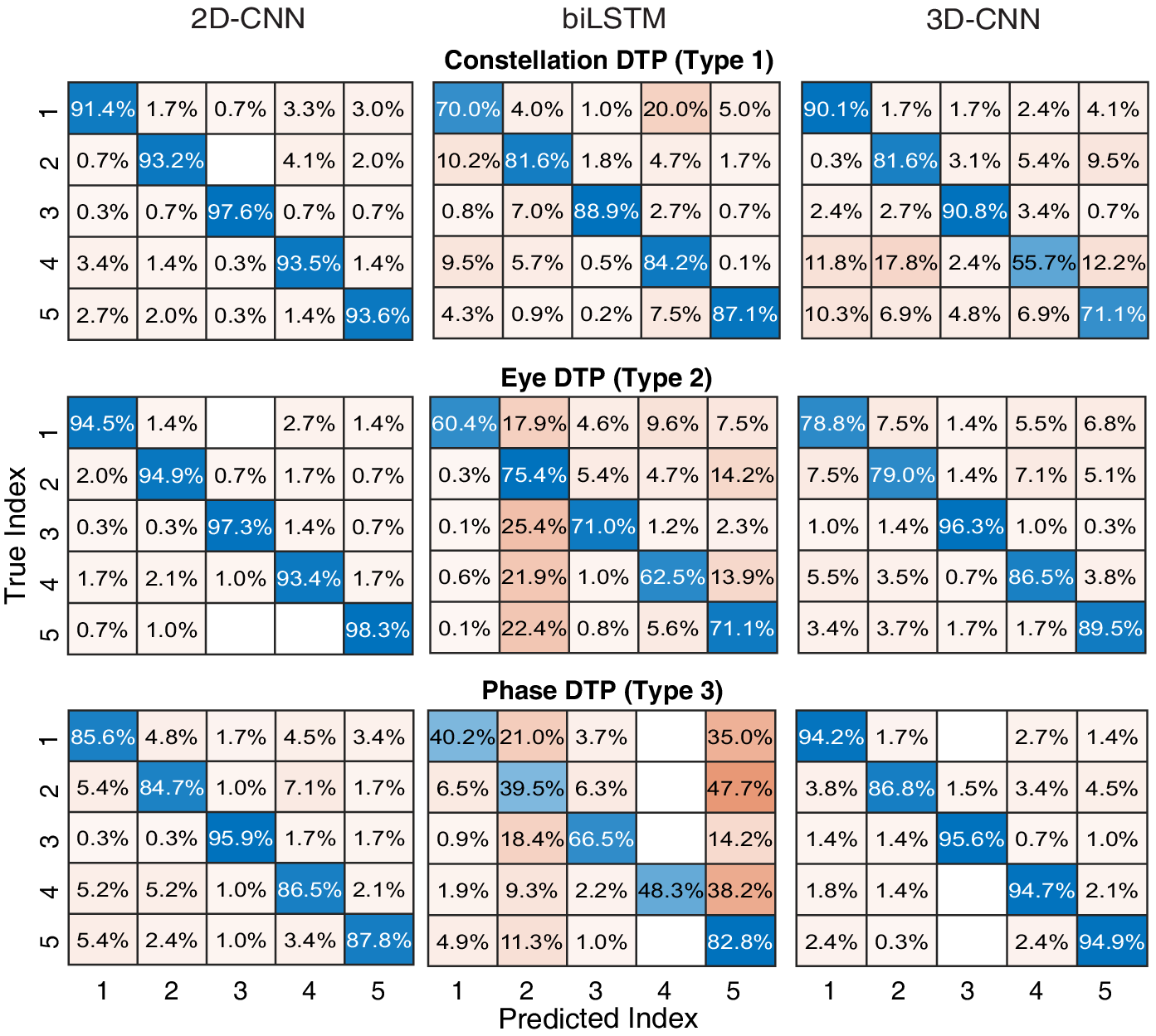}
\caption{Confusion matrices (row-normalized) for networks trained and evaluated using over-the-air 4-QAM samples, tested at SNR=10~dB.}
\label{fig:CM_4QAMwl}
\end{figure}

The confusion matrices for networks trained for the 4-QAM scheme at SNR of 10~dB are presented in Fig.~\ref{fig:CM_4QAMwl} to compare the performance of different possible combinations under the presence of the wireless channel. The highest accuracy across all three DTP types is again achieved by utilizing 2D-CNN. Meanwhile, the 2D-CNN+biLSTM architecture exhibits the worst overall accuracy among all three deep learning models. Being a form of Recurrent Neural Network (RNN), the biLSTM's inherent nature might limit its ability to normalize the distributions of feature maps across different time steps. This limitation can lead to poor performance when the inputs exhibit diverse distributions, such as affected by the presence of stochastic channels.

\subsection{Discussions}
When comparing the average performance across different modulation schemes, the BPSK modulation without additional phase shift yields the worst results. Given that the ideal BPSK modulates no data on its Q channel, the relatively simple inter-symbol transition might reduce the detectable fingerprints caused by physical layer impairments. Introducing an additional phase shift in the BPSK generates non-zero values for $x_\text{q}(t)$ in (\ref{eq:Ssync}), thereby increasing the presence of fingerprints caused by I and Q channel mismatch. Similar observations from the performance of other modulation schemes indicate that a more diverse inter-symbol transition can contribute to increased detectable fingerprints. However, isolating different symbols into distinct density centers might be challenging for DTP generated with a limited size. One straightforward solution is to increase the size of DTPs, but the trade-off between the increased resolution and extra computational overhead needs further investigation.

Regarding classification performance among the three proposed DTP types, the Type 1 DTP is found to be the most effective. A possible reason can be attributed to the focus on IQ imbalance as the sole source of RF fingerprints. Being mainly a modulation domain impairment, the distorted constellation due to device-specific IQ imbalance is often more distinguishable than the more complex changes observed in the eye diagram and the phase plot. Hence, testing how Type 2 and 3 DTPs react to other impairment sources could be valuable. 
Overall, as the targeted RF fingerprints are found to originate from device-dependent inherent characteristics, the proposed method embraces decent potential in its scalability if a suitable representation modality is used.

When comparing all three deep learning approaches employed, the two 3D networks show slightly worse performance than the 2D-CNN. It might be because the relatively short transmission duration limits the amount of potential temporal features that can be captured within 3D-DTPs. On the other hand, as observed in Fig.~\ref{fig:CM_16QAM2ms}, the 3D networks start to outperform the 2D-CNN when a longer source signal is introduced. The improved performance suggests that 3D networks require more temporal context to reach their full potential. It is worth noticing that the limited memory only allows us to iterate through limited hyperparameter settings during 3D network optimizations, especially for the resource-intensive 3D-CNN.  
When only comparing two of the 3D networks, the worse performance of 2D-CNN+biLSTM is potentially caused by the biLSTM module failing to efficiently identify varying input data distribution, especially with randomness from the wireless channel. Meanwhile, the 3D-CNN exhibits more consistent performance. This robustness is likely owing to the utilization of batch normalization layers, which allows the network to adjust feature map distribution. 
In addition, it is found that, when compared to both 2D and 3D CNNs, a trained 2D-CNN+biLSTM network requires more storage space and takes relatively longer to accommodate per prediction. 
The selection of classifiers indeed affects how efficiently RF fingerprints can be utilized for authentication. Thus, investigating the performance of different model architectures is valuable for enhancing the proposed method's scalability.

\section{Conclusion}
\label{sec:C}
In this work, we proposed a novel signal representation approach DTP to identify RF transmitters based on their physical layer properties. The presented data modalities exploit hardware impairment IQ imbalance solely as the clue for identifying different transmitters. Presented using three different modalities, the proposed data representation can be handled with either 2D or 3D neural networks for more flexibility. The experimental results show the best-performed DTP and deep learning pair can accurately classify five transmitting devices with an overall classification accuracy of up to 96.7\%. The results indicate the potential for the proposed approach to be applied, especially to IoT systems implemented using off-the-shelf devices, given that no structure modification is required. This work serves as the validation of our method. Indeed, we acknowledge a few limitations in the scope of this work that need to be addressed in future work. For instance, investigation under a more complex environment (e.g., with more transmitter devices, under the presence of different channel conditions) is crucial and will be addressed in future works.

\bibliographystyle{IEEEtran}
\bibliography{mylib.bib}

\end{document}